\documentclass{aa}
\usepackage{stfloats}
\usepackage{float}
\usepackage{color}
\usepackage{hyperref}
\usepackage{txfonts}
\usepackage{aas_macros}
\usepackage{natbib}
\usepackage{deluxetable}
\usepackage{lscape}
\usepackage{graphicx}

\begin{document}

   \title{Optical-NIR analysis of globular clusters in the IKN dwarf spheroidal: a complex star formation history}

%  \subtitle{}

   \author{ A. Tudorica\inst{1},
          I. Y. Georgiev\inst{2}
          \and
          A. L. Chies-Santos\inst{3,4}
          }
\offprints{tudorica@astro.uni-bonn.de} 

\institute{Argelander Institut f\"ur Astronomie der Universit\"at Bonn, Auf dem H\"ugel 71, D-53121 Bonn, Germany
\and Max-Plank-Institute f\"ur Astronomie, K\"onigsthul 17, 691717 Heidelberg, Germany
\and Departamento de Astronomia, Instituto de F\'isica, Universidade Federal do Rio Grande do Sul, Porto Alegre, R.S, Brazil
\and Instituto de Astronomia, Geof\'isica e Ci\^encias Atmosf\'ericas, Universidade de S\~ao Paulo, S\~ao Paulo, SP, Brazil}

\date{Received XX/XX/XXXX ; Accepted 04/06/2015}

%\abstract{}{}{}{}{} 

\abstract 
  % Context
{Age, metallicity and spatial distribution of globular clusters (GCs) provide a powerful tool to reconstruct major star-formation episodes in galaxies. IKN is a faint dwarf spheroidal (dSph) in the M\,81\,group of galaxies. It contains five old GCs, which makes it the galaxy with the highest known specific frequency ($S_N\!=\!126$).
}
  % Aims
{
We estimate the photometric age, metallicity and spatial distribution of the poorly studied IKN GCs. We search SDSS for GC candidates beyond the HST/ACS field of view, which covers half of IKN.
}
  % Methods
{To break the age-metallicity degeneracy in the $V\!-\!I$ colour we use WHT/LIRIS $K_S$-band photometry and derive photometric ages and metallicities by comparison with SSP models in the $V,I,Ks$ colour space.}
  % Results
{IKN GCs' $VIKs$ colours are consistent with old ages ($\geq\!8$\,Gyr) and a metallicity distribution with a higher mean than typical for such a dSph ([Fe/H$]\!\simeq\!-1.4_{-0.2}^{+0.6}$\,dex). Their photometric masses range ($0.5 <{\cal M_{\rm GC}}<4\times10^5M_\odot$) implies an unusually high mass ratio between GCs and field stars, of $10.6\%$. Mixture model analysis of the RGB field stars' metallicity suggests that 72\% of the stars may have formed together with the GCs.
Using the most massive GC--SFR relation we calculate a SFR of $\sim\!10M_\odot/$yr during its formation epoch. We note that the more massive GCs are closer to the galaxy photometric centre. IKN GCs also appear spatially aligned along a line close to the IKN major-axis and nearly orthogonal to the plane of spatial distribution of galaxies in the M\,81 group. We identify one new IKN GC candidate based on colour and PSF analysis of the SDSS data. 
}
  % Conclusions
{The evidence towards $i)$\,broad and high metallicity distribution of the field IKN RGB stars and its GCs, $ii)$\,high fraction and $iii)$\,spatial alignment of IKN GCs, supports a scenario for tidally triggered complex IKN's SFH in the context of interactions with galaxies in the M\,81 group.
}


\titlerunning{IKN SFH from optical/NIR photometry of globular clusters}
   
\authorrunning{A. \protect{Tudoric\u{a}}, I.Y. Georgiev, A.L. Chies-Santos}   
\maketitle
\section{Introduction}\label{Section:Intro}

Globular clusters (GCs) are compact stellar systems which typically form on a short time-scale, and as such, their stellar populations are composed of stars with similar age and metallicity. Therefore, the GCs' integrated light can be closely approximated by a single stellar population (SSP) with a  given age and metallicity. This makes GCs unique tracers of past \emph{major} star formation episodes in a galaxy's star formation history (SFH). They are often used to constrain galaxy formation models \cite[e.g.][and references therein]{Harris91,Ashman&Zepf92,Forbes97,Kissler-Patig00,Brodie06,Harris10, mg10, tonini13,Brodie14}. 

IKN is a faint, low surface brightness dwarf spheroidal (dSph) galaxy ($M_V\!=\!-11.5$\,mag, i.e. $L_V=3.4\times10^6L_\odot$). It is located in the outskirts of the nearby M81 group of galaxies \cite[$m\!-\!M\!=\!27.79\pm0.02$\,mag, 3.61\,Mpc,][]{Dalcanton09}, and at about 82\,kpc projected distance from M\,81. Recent analysis of Hubble Space Telescope (HST) imaging data with the Advanced Camera for Surveys (ACS) revealed that it has five GCs whose $V\!-\!I$ colours and half-light radii are consistent with being old and metal-poor \citep{Georgiev09}. The brightest of these GCs, IKN-05, was indeed spectroscopically confirmed to be a metal-poor GC \citep{Larsen14}. It is intriguing to find such a large number of old GCs in such low luminosity dSph. This makes it the galaxy with the highest observed number of old GCs per unit galaxy light, i.e. a specific frequency \citep{Harris&vdBergh81} of $S_N=126$ \citep{Georgiev10}. This value is significantly higher than for other galaxies of comparative luminosity as well as in more massive galaxies \cite[$S_N\leq20$,][]{Harris91,Bridges91,Villegas08,Peng08,Peng11,Harris09,Harris13,Georgiev10,Hargis&Rhode12}.

Such high $S_N$ therefore poses the puzzling question of what mechanisms could be responsible for the formation of the unusually large number of GCs that make up $\simeq\!13\%$ of IKN's luminosity? A plausible explanation could be offered by a complex formation history of the IKN, triggered by tidal interactions (tidal formation?) with more massive galaxies in the M\,81 group. Such a scenario is supported by the finding that IKN field stars have a broad metallicity distribution, which peaks at a too high metallicity for its luminosity, $[$Fe/H$]=-1.08\pm0.16$\,dex \citep{Lianou10}. Dwarf galaxies with similarly high metallicity for their luminosity have been also observed in the M\,81 group and were also suggested to have formed during galaxy interactions in the group from tidal debris \cite[e.g.][]{Croxall09}. Here we aim to test this scenario by studying the integrated properties of the IKN GCs. In addition, the high fraction of stars in GCs carries important information and implications for questions like how many stars formed in clusters in a given star formation epoch and how many clusters remain bound, i.e. implications for star cluster dissolution \cite[e.g.][]{Lada03,Bastian08,Goddard10,Kruijssen12,Larsen14}.

In this study we aim to improve the age and metallicity information for all GCs in the IKN dwarf spheroidal (dSph), by combining optical and near-infrared (NIR) photometry. 
We use existing optical \cite[HST/ACS,][]{Georgiev09} and new near-infrared $K_S$-band data from LIRIS on the William Herschel Telescope (WHT). 
The latter is very important to break the age metallicity degeneracy in the optical. This problem can be partially resolved with a combination of optical and NIR colour indices, as demonstrated by earlier work \cite[e.g.][ and refs therein]{Puzia02, Hempel03, Hempel07, Cantiello07, Chies-Santos11, Chies-Santos11a, Chies-Santos12, Georgiev12}. Additional complications in the optical domain come from the hot evolved stars (HB, blue stragglers) which contribute significantly to the integrated light, whereas in the NIR and for ages $\geq\!2$\,Gyr, the red giant stars completely dominate the total energy emission. In effect, NIR colours such as $I\!-\!K$ or $J\!-\!K$ measure the effective temperature of red giant branch stars, which in turn is very sensitive to metallicity, without a strong age dependence \citep{Worthey94}. An advantage of the optical-NIR technique, when compared to spectroscopy, is that for lesser amount of observing time, the properties of a statistically representative sample of GCs can be estimated. In galaxies beyond the Local Group resolving stars is very difficult with current ground-based telescopes. However, GCs are detectable and can be studied as far as the Centaurus, Hydra I and Coma galaxy clusters ($\approx 100$\, Mpc away) \citep[e.g.][]{Peng11}, thus providing a reliable and powerful tool to study  galaxy formation in the local universe. 
In addition, knowing about the number of GCs helps us shed light on galaxy formation theories. Therefore, we perform a search for new GC candidates in SDSS around IKN. The GCs are expected to differ from stellar point spread functions (PSFs) due to their more extended nature (effective radius of about 3\,pc) and thus would appear marginally resolved even in ground based imaging, such as the ones from SDSS. 

The outline of the paper is as follows: in Section \ref{Section:Observation and Data reduction} we describe the NIR-Optical data set and the reduction procedures. In the sections in \S\,\ref{Sect:Analysis_Discussion} we analyze and discuss the implications for the formation of IKN based on the photometric properties of the IKN GCs from derived photometric age, metallicity and mass (\S\,\ref{Section:Photometric_props}), SFR (\S\,\ref{Sect:PeakSFR}), mixture models for the metalicity distribution of IKN RGB stars (\S\,\ref{Sect:MDFs}) and GCs' spatial distributions (\S\,\ref{Sect:spatial_dist_center} and \S\,\ref{Section:GCs_alignment}). In \S\,\ref{Section:Conclusions} we summarize our results.

\section{Observations and data reduction}\label{Section:Observation and Data reduction}
\subsection{Observations}

The IKN dSph was observed in the $K_S$-band (2.15\,$\mu$m) with LIRIS (Long-slit Intermediate Resolution Infrared Spectrograph), mounted on the 4.2\,m William Herschel Telescope (WHT), during two consecutive nights (15 and 16 March 2009).
LIRIS provides a field of view of $4.27 \times 4.27$\,arcmin$^2$, thus covering the entire galaxy and its immediate surroundings. A plate scale of $0.25$ $\arcsec$/pixel ($\approx4.5$\,pc/pixel at distance of 3.61\,Mpc to IKN) means that its globular clusters will be observed as unresolved extended sources. 

The $K_S$-band observations (see \citealt{Chies-Santos11a}) were performed in a dither pattern of five consecutive exposures at five different positions in a cross-like arrangement with offsets between 10\arcsec and 20\arcsec. The exposure time of a single image (DIT) was kept short (15\,seconds) to deal more easily with the high sky background level in the near-IR. This observing strategy facilitated the sky subtraction reduction step (see Section \ref{Section:NIR_data_reduction}).

The final exposure time of all combined images sums up to 5505\,seconds, which allowed us to perform photometry on the faintest IKN GC ($V=21.2$\,mag, i.e. $M_V=-6.65\pm0.06$\,mag) for a distance modulus adopted here of $(m\!-\!M)=27.79\pm0.03$\,mag. The observation log is summarized in Table \ref{Table:observations}. A total of 6 photometric standard stars for photometric calibration were observed in each night. 

\begin{table}[ht]
  \centering
    \caption{Observations log and night conditions.}\label{Table:observations}
\begin{tabular}{c|cccc}
    \hline\hline
    Night      &      FWHM          & N       & Total exposure time  \\ 
    yyyy-mm-dd &     $\arcsec$       & images  & seconds              \\ \hline
    2009-03-15 &   $\approx$ 1.0    & 76      &  1140                \\
    2009-03-16 &   $\approx$ 0.9    & 291     &  4365                \\
    \hline\hline
  \end{tabular}
\end{table}

The V and I photometry of the IKN GCs \citep{Georgiev09} used in this study is based on 
the HST/ACS archival data program SNAP-9771 (PI: I. Karachentsev). The non-dithered $2\times$\,600\,s $F606W$ and $2\times450$\,s $F814W$ exposures were designed to reach the Tip of the Red Giant Branch (TRGB) for distance measurement \citep{Karachentsev07}.

To extend the search for GCs around IKN in regions not covered by the HST/ACS observations, we used the Sloan Digital Sky Survey (SDSS) $ugriz$ DR10 archival images\footnote{http://www.sdss.org/dr10/} \citep{Abazajian09}. SDSS is a large scale survey with the $2.5m$ Apache Point Observatory telescope covering a quarter of the sky. SDSS reaches a resolution of $0.396\arcsec$/pixel with an r-band seeing of $1.55\arcsec$ and limiting magnitudes sufficiently deep to enable a search for GC candidates. The pixel scale corresponds to a projected spatial resolution at the IKN distance of 7\,pc/pixel and the corresponding seeing value of $1.55\arcsec$ to a Gaussian FWHM of $13.5\,pc$. The $ugriz$ data allows us to extend the wavelength coverage for the already known clusters and search for new candidates (see details in Section \ref{Section:SDSS_candidates}). All magnitudes are in the Vega system.

\subsection{Data reduction and photometry}\label{Section:Data}
\subsubsection{LIRIS data reduction}\label{Section:NIR_data_reduction}

\begin{figure*}
\resizebox{1.0\hsize}{!}{\includegraphics[keepaspectratio, angle=0]{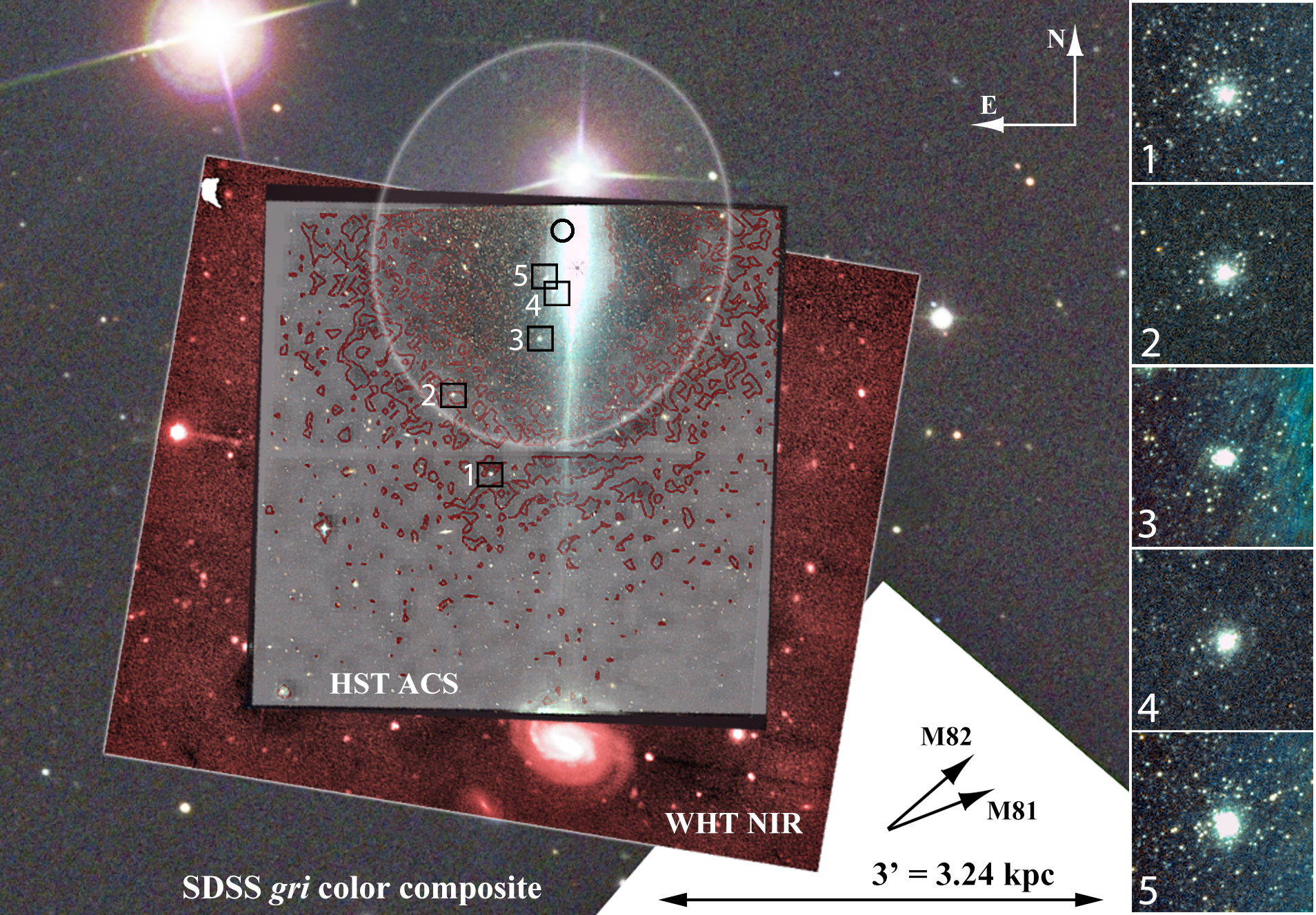}}\hfill
\caption{SDSS $gri$ colour composite mosaic of the IKN dSph field and its five known GCs. These are indicated with squares in the colour image and shown magnified on the right from the HST/ACS image data. The labels indicate the field of view for the archival HST/ACS $F606W, F814W$ and our WHT/LIRIS $K_S$-band image. The light-colour stellar density contour map is adapted from  Figure\,6 in \cite{Lianou10}. The ellipse indicates the HyperLEDA values for the IKN diameter (2.69\arcmin\,=\,2.91\,kpc) and ellipticity (0.18). For reference, with the arrows we indicate the direction to the group dominant galaxies M\,81 and M\,82.}
\label{Figure:IKN_HST_Kband}
\end{figure*}

The NIR imaging reduction was performed within the LIRIS IRAF package LIRISDR\footnote{Package written by Jose Acosta Pulido to manage data produced by the LIRIS instrument - www.iac.es/galeria/jap/lirisdr/LIRIS\_DATA\_REDUCTION.html}. For each night the data was first corrected for the pixel mapping anomaly with the {\em lcpixmap} IRAF task, an effect produced by the reconstruction of the two-dimensional array which misplaces some of the pixels, especially in the lower left quadrant. 

The images were flat-field corrected and sky subtracted using the $ldedither$ LIRISDR task for each night. One of the first corrections performed is the flat-fielding, which must correct not only for large scale gradients in the image but also for pixel-to-pixel gain variations. The dome flat fields were obtained by subtracting dark frames from bright frames of the same exposure time in order to suppress the possible thermal contamination of the telescope environment and then coadded and normalized for each night. The sky subtraction step is the most important step in the data reduction of NIR data as the brightness and structure of the NIR sky varies on a timescale of minutes and its contribution must be individually subtracted for each pixel. The general idea is to obtain the median frame for a set of images, discarding the values of the pixels that have received flux from celestial objects and then subtract the median frame from the science frames. The $ldedither$ procedure creates sky frames from sets of five consecutive (dithered) exposures and generates a sky subtracted image for each frame before coaddition. 

  We aligned all non-registered, but sky-subtracted output images from $ldedither$ using a custom IRAF procedure based on $geomap$, $geotran$ and $imcombine$ IRAF tasks. We used a sigma-clipping coaddition algorithm, with a statistics region used to estimate the images zero offsets and the images are scaled by their exposure time. Figure\,\ref{Figure:IKN_HST_Kband} shows the final overlap region of the LIRIS coadded image with the HST and SDSS DR10 data. The standard stars data were reduced using the same IRAF and LIRISDR procedures as for the IKN images. The more photometric night was March 16, to which all images were subsequently aligned and registered. The total exposure time in the final coadded image is 5505s.

\subsubsection{Ks-band photometry of IKN globular clusters}\label{Section:GC_Kband_phot}

Aperture magnitudes of the clusters, standard and field stars were measured with tasks in the IRAF DAOPHOT package. We performed a curve of growth analysis with $r=2,3,4,...,40$ pixels aperture radius for the GCs and high signal to noise isolated field stars in the images, subsequently choosing a photometry radius of 12 pixels. This is about three times the average full width half maximum of stars and unresolved sources in the final co-added image. The curve of growth analysis from isolated stars allowed us to derive an aperture correction of 0.04 magnitudes. PSF photometry was not possible due to the low number of good PSF stars necessary to build a reliable PSF model.

The photometry is calibrated assuming a transformation equation of the form below, where the coefficients are derived from a least squares fit to the magnitudes of the photometric standards:

\begin{equation}\label{Equation:transformation_eq}
K_S = k_s + z_k + p_k\times x_{k_{eff}} + c \times (V-k_{S}).\\ 
\end{equation}
$K_S$ and $k_s$ are the standard and instrumental magnitudes, $p_k\!=\!- 0.07 \pm 0.02$ is the atmospheric extinction coefficient \citep{Chies-Santos11a},  $x_{k_{eff}}$ is the effective airmass and $V$ is the $V$-band magnitude of the standard stars adopted from \cite{Jenkner90}. Due to the low number of $K$-band standards with reliable published $V$-band magnitudes, the value of the colour term $c$ was adopted to be $0.02$, as found in earlier studies \citep{Goudfrooij01,Georgiev12}. The zero point was evaluated from the fit to the standard stars to be $23.01\pm0.02$\,mag, not too different from the value found by \citep{Chies-Santos11a} of $23.07\pm0.01$\,mag for another set of observations.
	
The final calibrated $K_S$-band magnitudes of the clusters were estimated by using their $V$-band magnitude published in \cite{Georgiev09}, iterating Equation \ref{Equation:transformation_eq} until the $K_S$ magnitude converged. The $V$ and $K_S$ magnitudes were also corrected for foreground galactic extinction with values for the total absorption of $A_V = 0.22$\,mag, $A_{Ks}=0.02$\,mag taken based on \cite{Schlafly11} re-calibration of the \cite{Schlegel98} dust maps assuming the \cite{Fitzpatrick99} reddening law with $R_V\!=\!3.1$.
	
The foreground reddening corrected IKN GCs photometry is tabulated in Table\,\ref{Table:gc_properties1}.
	
\subsubsection{Search for GC candidates in SDSS DR10}\label{Section:SDSS_candidates}	

Since HST/ACS field of view covers half of IKN, we utilize the wide spatial and spectral coverage of SDSS to search for GC candidates. IKN dSph is not entirely covered by the HST/ACS field of view and the stellar density map \citep{Lianou10} indicates that the center of the IKN field stars is approximately at the location of the bright foreground star near the edge of the field of view (see Fig.\,\ref{Figure:IKN_HST_Kband}, also \S\,\ref{Sect:spatial_dist_center}). Therefore,  given the apparent distribution of GCs on one side of IKN additional GCs may be located outside the HST/ACS image. Searching for GC candidates beyond the optical extent of IKN is also motivated by the discovery of very distant GCs around the dwarf irregular galaxy NGC\,6822 located out to 11\,kpc from the galaxy centre \citep{Hwang11}. These GCs have similar luminosities, sizes and spatial distributions as the known IKN GCs (see \S\,\ref{Section:Conclusions} and \S\,\ref{Section:GCs_alignment}).

To search for GC candidates in IKN we make use of their nature as extended objects. HST/ACS derived half-light radii for the known GCs are in the range $ 1.96\pm 0.16$\,pc to $14.81\pm 0.83$\,pc \citep{Georgiev09}. The r-band seeing of the SDSS images used is 1.55\arcsec, which corresponds to a projected spatial scale at the IKN distance of 27\,pc. This implies that at least the larger half-light radii GCs will have a PSF which differs from a point-source PSF.

Although the model and PSF magnitudes in the SDSS photometric catalogues are quite reliable, we perform our own PSF photometry on archival SDSS images in order to evaluate the difference between stellar PSF and GC profiles and search for new cluster candidates. The additional reason to perform our own PSF analysis is because IKN falls on two SDSS stripes, which requires a separate PSF analysis. We find the difference between the PSFs of the two stripes to be well within the measurement uncertainty. We selected isolated, high S/N stars to build a PSF model and performed a PSF photometry on all detected objects with the IRAF's DAOPHOT package procedures ($psf$ and $alltstar$). Instrumental PSF magnitudes were calibrated to standard SDSS magnitudes by using the SDSS PSF magnitudes of the same stars. The final SDSS photometry of the IKN GCs is tabulated in Table\,\ref{Table:gc_properties1}.

\begin{figure}[ht]
\includegraphics[width=0.5\textwidth, viewport=10 15 350 250]{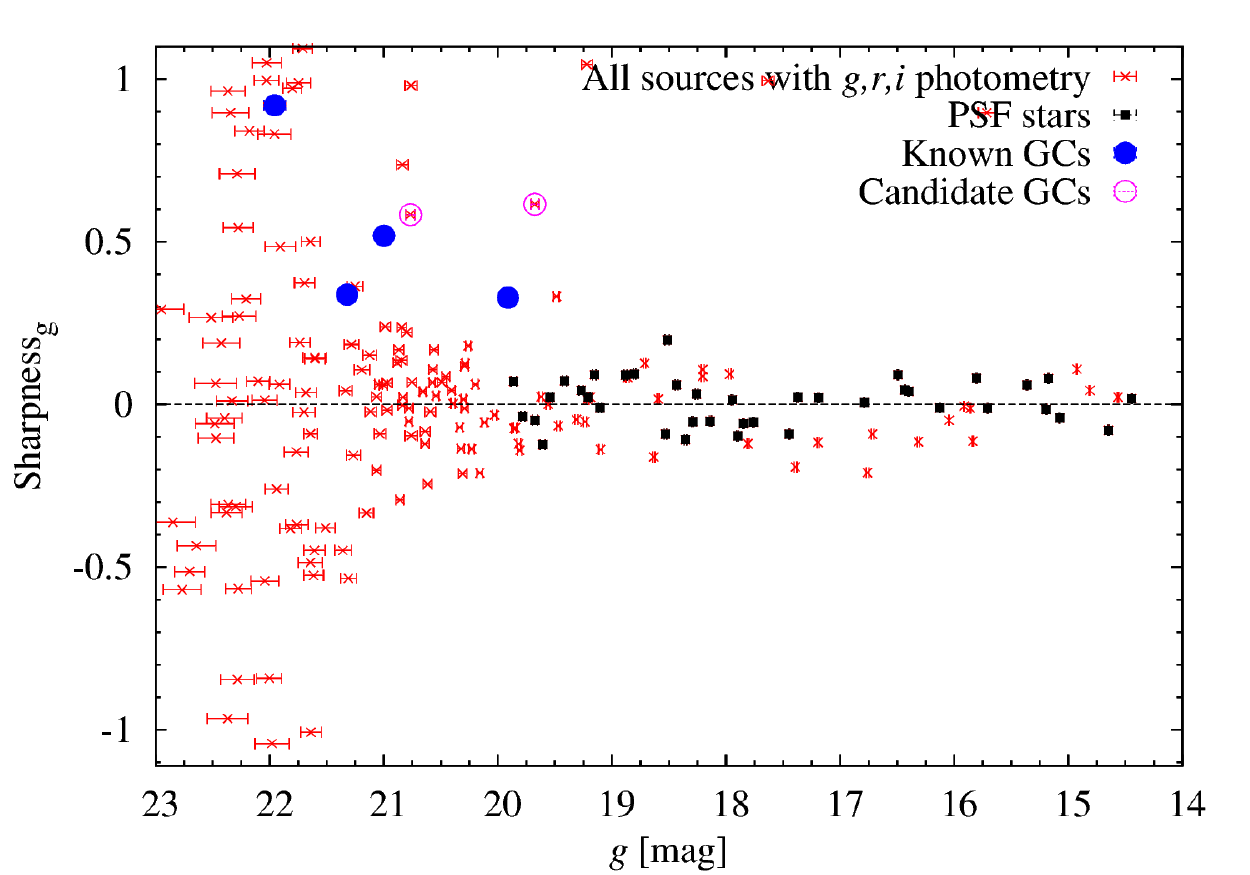}
\caption{Selectinon of GC candidates based on sharpness value. Shown is a $g-$band example of the sharpness value versus magnitude for all sources with PSF photometry (small crosses), the known IKN GCs (large solid, blue circles) and the SDSS GC candidates (open circles). The PSF stars are shown with small solid symbols (dark circles).}\label{Fig: Sharp vs Mag}
\end{figure}

Due to the fact that the clusters were not detected or with too low S/N in the $u$-band images, this band was not used in the subsequent analysis. We used the sharpness value returned by $allstar$ during the PSF fitting, to search for objects more extended than the stellar PSF with similar $sharp-$value as the known GCs. The $allstar$ sharpness parameter is defined as the difference between the squares of the width of the PSF model and the width of the object. That is, sources with light profile similar to the stellar PSF will have $sharp-$value close to 0, while more extended sources will have a $sharp-$value greater than 0. The extended nature of the candidates can be seen in Figure\,\ref{Fig: Sharp vs Mag}, where the sharpness values for PSF stars, GCs and all other objects are plotted as a function of their $g$-band magnitude. The second selection criteria was based on colours, with the candidates having to be in the immediate vicinity of the GC's in the griz colour space (within 2\,$\sigma$ from the average GC colours).
	
One of the new GC candidates is very close to the bright foreground star (BD69+557), which prevents reliable photometry. Therefore we do not consider it for further analysis. 

\section{Analysis and Discussion}\label{Sect:Analysis_Discussion}

\subsection{GCs' colours, photometric age, mass and metallicity}\label{Section:Photometric_props}
     
To gain insight on the formation epochs of the IKN GCs and the evolutionary path of this faint dSph, we discuss the stellar population age and metallicity of the IKN GCs based on their optical-NIR integrated colour distributions. As mentioned earlier, we take advantage of the good age and metallicity resolution provided by the optical-NIR colour-colour indices. 
     
\begin{figure}
\includegraphics[width=0.5\textwidth, bb=20 20 350 245]{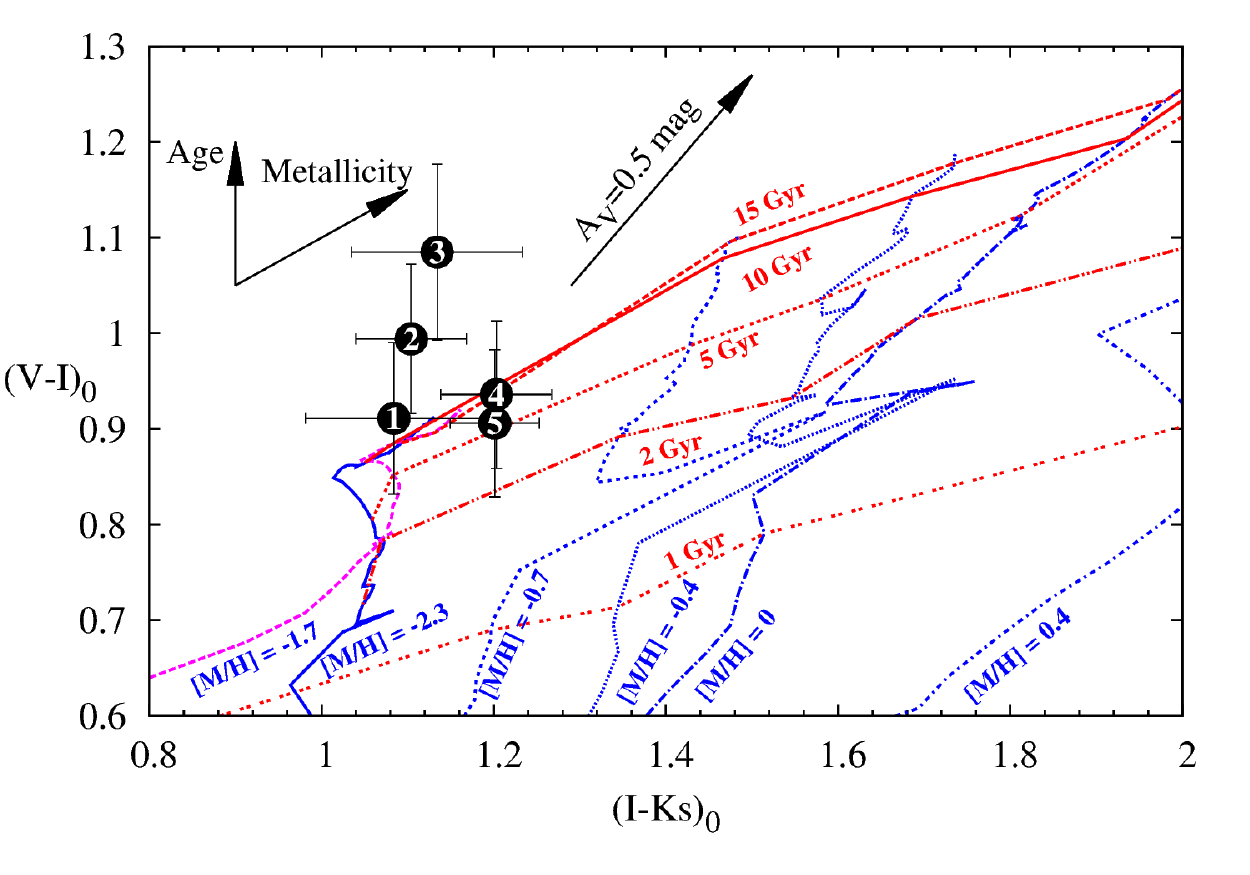}
\caption[Optical\,-\,NIR colour-colour $(V\!-\!I)_0$\,v\,$(I\!-\!Ks)_0$ plot] {Foreground redenning corrected optical\,-\,NIR colour-colour plot, $(V\!-\!I)_0$\,v\,$(I\!-\!Ks)_0$ of the IKN globular clusters (solid labeled circles). The cluster colours are compared to \cite{Bruzual03} SSP model isometallicity tracks (M/H labeled lines) and isochrones (age labeled lines). The top left arrows show the direction of increasing age and metallicity and the arrow to the right shows a reddening vector of $A_V\!=\!0.5$\,mag.}
\label{Figure:CCD}
\end{figure}

Figure\,\ref{Figure:CCD} shows the $(V\!-\!I)_0$\,vs\,$(I\!-\!Ks)_0$ colour-colour distribution of all GCs in the IKN dSph, compared to \cite{Bruzual03} SSP models for a canonical  \citep{Kroupa01} IMF, and a range in age and metallicity as indicated by the labels.
The distribution of the GCs' colours in Figure\,\ref{Figure:CCD} suggests that all of them are likely old and metal-poor [M/H]$\leq -0.7$ dex. This is consistent with the metal-poor and old age inferred for the brightest IKN GC-5 from recent spectroscopic analysis \citep{Larsen14}. However, given the photometric uncertainties, ages as young as 2-5\,Gyr cannot be ruled out. In general, although with lesser certainty, the photometric metallicity spread of the IKN GCs is consistent with that of the IKN field stars, [Fe/H]$\leq -1.0$ dex \citep{Lianou10}.
We have also compared the IKN GCs with the GC systems of the luminous elliptical galaxies M87 and M60 from \cite{Chies-Santos11} in a gzk colour-colour diagram. IKN GCs 4 and 5 fall on the same loci as M87/M60 GCs in the $gzk$ parameter space, implying old ages. The other IKN GCs do not have $gzk$ photometry.
     
To derive photometric ages and metallicities for the GCs in our sample we perform a minimization interpolation between the age and metallicity in the \cite{Bruzual03} SSP model grid. These are estimated with the average value between the two closest values weighted by the distance to the respective model and the photometric uncertainty. This method enables us to derive upper and lower errors for the photometric ages and metallicities. The validity of this approach has been discussed in detail for the M\,31 and MW GC colours with \cite{Bruzual03} SSP tracks for old ages \citep[e.g.][]{Puzia02,Georgiev12}. GCs falling outside of the model grid have age and metallicity derived from the nearest age and metallicity model tracks. A comparison of this technique for different SSP model tracks is presented in \cite{Georgiev12}. Errors due to the choice of the SSP model are included in our measured values and amount to about 10\% at given age and metallicity. Thanks to the combination of optical-NIR magnitudes, uncertainty in derived cluster mass due to stochastic effects are below the 1-5\% level for \emph{old and massive GCs, $\gtrsim10^5M_\odot$} \citep{Fouesneau10,Fouesneau12}. In Table\,\ref{Table:gc_age_Z_Mass} the measured values for photometric age, metallicity and mass of the IKN GCs are tabulated. 
\begin{table}[H]
\centering    
\caption[Estimates for the IKN GCs properties]{Photometric metallicity, age and mass of IKN GCs.\\[-.2cm]}
    \label{Table:gc_age_Z_Mass}
\begin{tabular}{ccccccccc}
    \hline\hline%\\[-.2cm]
  ID   & Z & [Fe/H] &            Age                 &             Mass                          \\   
       &  [Z$_\odot$]            &  [dex]  &            [Gyr]               &       [$10^5 \, M_{\odot}$]                     \\[.1cm]
\hline\\[-.2cm]
IKN-01 &  $0.024^{+0.017 }_{-0.015}$   &  $-1.833_{-0.423}^{+0.235}$   &  $14.77^{+1.14}_{-1.30}$    & $0.771^{+0.069}_{-0.075}$ \\
IKN-02 &  $0.051^{+0.063 }_{-0.037}$   &  $-1.506_{-0.558}^{+0.353}$   &  $15.51^{+3.49}_{-6.18}$    & $1.261^{+0.273}_{-0.467}$ \\
IKN-03 &  $0.078^{+0.064 }_{-0.050}$   &  $-1.321_{-0.442}^{+0.263}$   &  $13.21^{+4.66}_{-5.99}$    & $0.779^{+0.235}_{-0.299}$ \\
IKN-04 &  $0.065^{+0.075 }_{-0.047}$   &  $-1.401_{-0.554}^{+0.337}$   &  $14.19^{+4.50}_{-7.24}$    & $1.478^{+0.433}_{-0.689}$ \\
IKN-05 &  $0.069^{+0.080 }_{-0.051}$   &  $-1.375_{-0.580}^{+0.338}$   &  $13.80^{+4.90}_{-7.74}$    & $3.824^{+1.247}_{-1.952}$ \\[.2cm]
\hline
  \end{tabular}
	\end{table}

\begin{figure}[H]
\includegraphics[width=0.5\textwidth, bb=10 10 340 280]{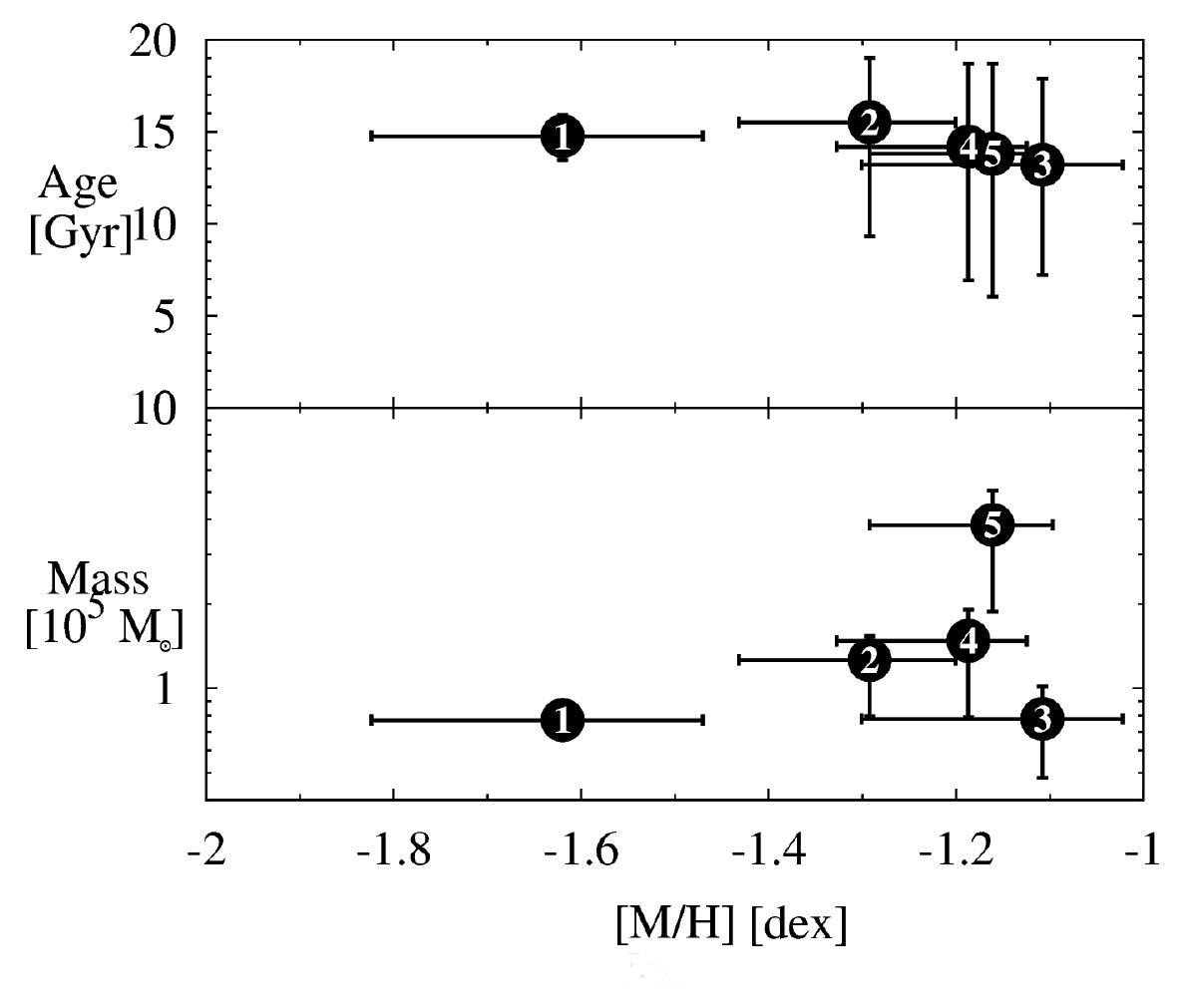}
\caption{Mass-metallicity (top) and age-metallicity (bottom) distributions of the IKN GCs (solid labeled circles) derived from their $VIK$ magnitudes.}
\label{Figure:MassZ}
\end{figure}

To convert from colour-magnitude to metallicity-mass distributions, we used the SSP model $M/L_V$ values based on the GC's colours to convert from luminosities to masses (for the adopted distance to IKN, see \S\,\ref{Section:Intro}). The resulting mass-metallicity distribution is shown in Figure \ref{Figure:MassZ}. The absolute values for the metallicities of the IKN GCs range from -1.6\,dex to -1.1\,dex and are relatively metal-poor, and are also consistent with metallicity values of the IKN foreground stars \citep{Lianou10}.
Figure\,\ref{Figure:MassZ} shows that the photometric GC ages are older than about 10\,Gyr and with a narrow spread. This suggests that the early formation of all clusters occurred within a relatively short time scale, which is typically the case for the old GCs in other galaxies, such as in the Milky Way or the old GCs in the LMC (see more detailed discussion of IKN RGB and GCs' metallicity distributions in \S\,\ref{Sect:MDFs}).

\subsection{IKN peak SFR from its most massive GC}\label{Sect:PeakSFR}

It has been demonstrated that there is an observational relation between the galaxy star formation rate (SFR) and the luminosity of its most massive cluster at a given epoch \citep{Larsen00,Weidner04,Maschberger07,Bastian08}. Here we derive the IKN peak SFR from the photometric mass of the IKN most massive GC identically to \cite{Georgiev12}. Briefly, this included a correction for globular cluster mass loss due to stellar evolution using the \cite{Bruzual03} SSP $M/L_V$ ratios at 10 Myr. We have not accounted for mass loss from tidal shocking, which however, could be expected to be small for such a low mass dwarf. Because the suggested age spread among the IKN GCs is very small (cf Fig.\,\ref{Figure:MassZ}), the IKN peak SFR is calculated from its most massive cluster. For reference, we show 
\begin{figure}[h]

\includegraphics[width=0.5\textwidth, bb=10 10 360 240]{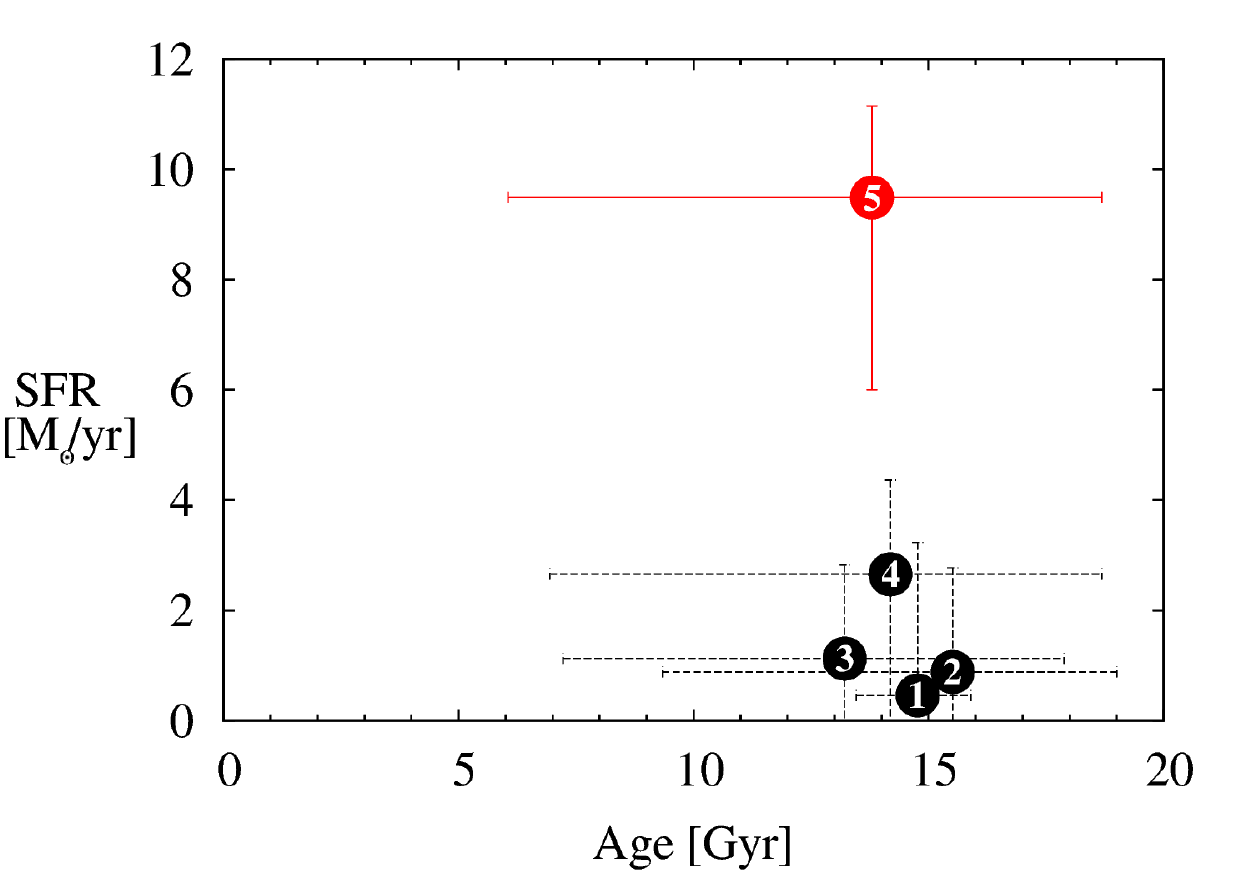}
\caption{Star formation rate calculated for each IKN GC against cluster age. The most massive GC (GC-5) yields a SFR  of $10 \, M_{\odot}/year$ at the time of GCs formation.}
\label{Figure:AgeSFR}
\end{figure}
in Figure \ref{Figure:AgeSFR} the age against the SFR derived for each GC. The upper limit estimate for the peak star formation is of $10 \, M_{\odot}/year$ and occurred around 14 Gyr ago. The most massive GC provides a lower limit of the SFR at  $14.29^{+1.01}_{-1.21}$ Gyr,   (the errors are propagated from the GCs mass uncertainties). The duration of the star formation burst cannot be estimated with high precision without good knowledge the GCs ages. However, given the low mass of the IKN and assuming a constant SFR, we can constrain the maximum length of the burst to a only a few (ten) million years. This indicates that IKN went through a bursty star formation episode at the epoch of the GCs formation.
          
Due to the very small number of globular clusters, one must keep in mind that statistical effects in the GC mass distribution might play a significant role in the SFR that we find for IKN.

The new GC candidates are either outside the HST and WHT field of view or within the region rendered unusable for analysis by the bright star. Thus, the candidates could not be confirmed visually on the HST image and that the NIR data is not available for breaking the age-metallicity degeneracy. 

\subsection{Metallicity distributions of GCs and field RGB stars - implications for IKN SFH}\label{Sect:MDFs}

The comparison between the metallicity distributions of the GCs and IKN's field stellar population and using the knowledge of the galaxy SFR, can provide important insights about how much of the initial star formation took place in star clusters.

In order to be able to directly compare the metallicity distribution function (MDFs) of the IKN GCs with that of the RGB field stars we convert their total metal content ([M/H], derived in \S\,\ref{Section:Photometric_props} from SSP models) into [Fe/H]. For that we use the \cite{Salaris93} relation between [M/H] and [Fe/H] (their eq.\,3), where the factor $f_\alpha=10^{\alpha}$ with $\alpha=0.3$\,dex, accounts for the typical high $\alpha$-element abundance (i.e. low Fe content) observed in GCs \cite[cf eq.\,1 in ][]{Ferraro99}. The converted [M/H] to [Fe/H] metallicity values for the IKN GCs are given in Table\,\ref{Table:gc_age_Z_Mass}. 

\begin{figure}
\includegraphics[width=.5\textwidth, bb = 10 10 350 340]{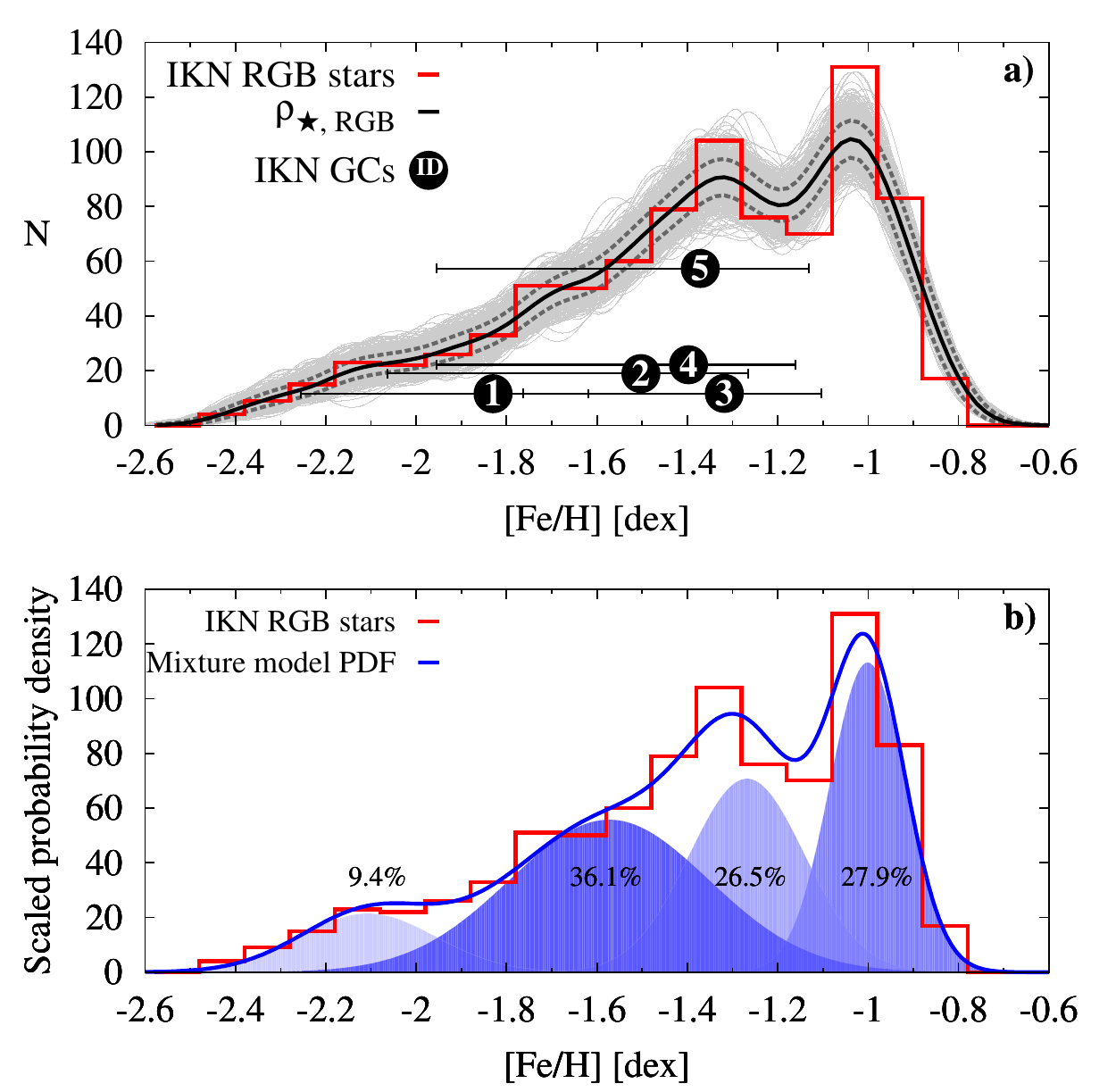}

\caption{{\bf Top:} Metallicity distribution of IKN field RGB stars (solid, red line histogram) and IKN GCs (labeled solid circles). For viewing purposes, the vertical position of the GCs is scaled according to their mass. Thick solid (black) and dashed (grey) curves are the scaled probability density estimate ($\rho_{\star,\rm RGB}$) and its $1\sigma$ estimated from the model PDFs (thin gray lines, details in \S\,\ref{Sect:MDFs}). {\bf Bottom:} Mixture model of the IKN field RGB stars metallicity distribution. The best fit maximum likelihood mixture model probability density distribution  is shown with the solid blue curve and its components by filled curve Gaussians. For comparison we show again the metallicity distribution of IKN field RGB stars (solid red line histogram). The percent fraction of each fitted Gaussian is given in the legend and Table\,\ref{Table:mixmodel_results}.
\label{Figure:fe_hist}}
\end{figure}

Figure\,\ref{Figure:fe_hist} shows the direct comparison between the MDFs of RGB field stars and IKN GCs. 
For viewing purposes only, the y-axis values of the GCs are obtained by multiplying their photometric mass, ${\cal M}_{\rm GC}$, (cf Table\,\ref{Table:gc_age_Z_Mass}) by $1.5\times10^{-4}$. \citeauthor{Lianou10} kindly provided us with the photometry of the IKN field RGB stars, which MDF is shown with a solid line histogram in Figure\,\ref{Figure:fe_hist}. \cite{Lianou10} have already reported that IKN has a broad MDF which peaks at higher [Fe/H] than for such low-luminosity dwarf galaxy. This suggests that IKN has likely experienced a complex SFH with probably more than one burst of star formation.

In order to assess the significance and the likelihood for a multimodal as opposed to a unimodal MDF and compare the fraction of stars in GCs, we performed a more detailed statistical analysis with packages in R\footnote{R is a free software environment for statistical computing.\\ \href{http://www.r-project.org/}{http://www.r-project.org/}}. First, we estimated the [Fe/H] probability density function (PDF) of the RGB stars using an Epanechnikov kernel density estimator. We used this PDF as a prior and the [Fe/H] uncertainty of the RGBs to generate a large number of models (1000). The different model realizations are shown with thin (grey) lines in Figure\,\ref{Figure:fe_hist}. The joint PDF of all the models, as a function of [Fe/H], gave the final PDF of the RGB stars [Fe/H] distribution. We show this in Figure\,\ref{Figure:fe_hist} with solid (blue) curve and its corresponding one sigma intervals are represented by dashed (grey) lines. This test confirms that there is a dip in the IKN field RGB's MDF at the $1\sigma$ level.

In order to further test the significance of the $1\sigma$ dip in the IKN MDF, we perform an additional mixture model statistical analysis to test whether the MDF is consistent with one or multiple underlying distributions. This particular test will allow us to assess whether the observed MDF can be represented by the combination of more than one Gaussian distribution. This will allow to derive the fraction of the stellar population which falls into the same metallicity range as the old globular clusters, i.e. within the same SF epoch. For that we used routines in the {\sc mixdist} and {\sc mixtools} R packages with which we fit a mixture Gaussian distribution models to the data by the method of maximum likelihood minimization. We performed a mixture models with one, two, three, four, five and six Gaussian components. The four Gaussian model was the preferred solution with considerably the best statistics (smallest $\chi^2$ and highest $p-$value) of all models. Models with higher number of Gaussians give a marginally better $\chi^2$, but at the same time give the worst $p-$value. This is because all components are simultaneously fit and likelihoods maximized. We can thus conclude from our statistical test that more than one component is needed in order to reproduce the MDF of the IKN RGB stars. We also note that only the four component model gave a good representation of the metal poor peak at [Fe/H$]\simeq-2.2$\,dex. It is beyond the scope of this paper to further investigate exactly how many components, i.e. episodes of star formation could have been responsible for the present day IKN field RGB's MDF. The values for each of the four fitted components are given in Table\,\ref{Table:mixmodel_results}. 
\begin{table}
\caption{Coefficients of the different Gaussian mixture model components. Column (1) gives the fraction of RGB stars in each of the components; column (2) and (3) give the Gaussians $\mu$ and $\sigma$ with their uncertainties.
\label{Table:mixmodel_results}}
\begin{tabular}{ccc}
\hline\hline
Population fraction   &  $\mu$ &  $\sigma$ \\
$[\%$]		&		[dex]		&		[dex]	\\
(1)		&		(2)			&		(3)	\\
\hline
$9.45\pm5.1$	 &	 $-2.11\pm0.084$  &	   $0.1497\pm0.039$ \\
$36.09\pm16.3$	 &	 $-1.57\pm0.129$  &	   $0.2199\pm0.067$ \\
$26.55\pm23.4$	 &	 $-1.27\pm0.041$  &	   $0.1276\pm0.077$ \\
$27.91\pm9.0$	 &	 $-1.00\pm0.023$  &	   $0.0837\pm0.011$ \\
\hline\hline
\end{tabular}
\end{table}

Our mixture model fitting suggests that nearly a third (about 28\%) of the IKN field stars are in the most metal-rich component, while the rest 72\% of the metal poorer RGB stars \emph{could} have formed in more than one SF event. The uncertainties in the parameters of the three metal-poor components (see values in Table\,\ref{Table:mixmodel_results}) prevent us from a more firm and detailed discussion whether and how many episodes of SF IKN could have went through. In addition, whether a Gaussian or another function would be a better description is also beyond the scope of this paper. For the following discussion we will consider that up to 72\% of the IKN star formation occurred simultaneously with the epoch at which the GCs formed. This stems from the distribution of the IKN GCs in Figure\,\ref{Figure:fe_hist}\,a) compared to the field stellar component of IKN, which fall in the metal poorer component.

We can first calculate the fraction of mass in clusters to total galaxy stellar mass. Summing up cluster masses in Table\,\ref{Table:gc_age_Z_Mass} gives ${\cal M}_{\rm GC,tot}=8.14\times10^5M_\odot$. For a total IKN luminosity of $M_V=-11.2$\,mag and a $M/L_V=3$ for [Fe/H$]=-1.0$\,dex and an age of 12\,Gyr, the IKN mass is thus ${\cal M}_{\star,\rm IKN}=7.7\times10^6M_\odot$. Therefore we obtain a mass fraction of $f_{{\cal M,}\rm GC}=0.106$, i.e. 10.6\% of the IKN stellar mass is in its old GCs. This is unusually high compared to the fraction of light or mass in GCs in more massive early- and late-type galaxies \cite[e.g.][]{Peng08,Georgiev10,Harris13}. If we consider our results from the mixture models above, i.e. that probably about 72\% of the stars in IKN have formed in the same star formation episode as the old GCs, we will arrive at an even larger fraction of about 14\%.

It is interesting to ask how the fraction $f_{{\cal M,}\rm GC}$ relates to the actual fraction of stars that formed in clusters, $\Gamma=$\,CFR/SFR, introduced by \cite{Bastian08}.  We can use the SFR =10 $M_\odot/$yr value obtained in \S\,\ref{Sect:PeakSFR} and calculate the surface SFR, $\Sigma_{\rm SFR}$. We adopt a radius equal to the optical extent of IKN, $r=1.5$\,kpc. Thus, $\Sigma_{\rm SFR}=1.4 M_\odot/$\,yr\,/\,kpc. This value for $\Sigma_{\rm SFR}$ suggests that $\Gamma$ should be in the 
range $\Gamma=10-30\%$ based on the $\Gamma - \Sigma_{\rm SFR}$ relation recently investigated for a wide range of environments \cite[e.g.][]{Bastian08,Goddard10,Kruijssen12,Mora14}. This further confirms that IKN had a very intense SFH that lead to high CFRs and to the very high present day fraction of old GCs. 
     
\subsection{Spatial distributions of the IKN GCs}\label{Sect:spatial_dist_center}

Investigating whether the mass of the IKN GCs varies as a function of the galactocentric distance may hold important information about cluster formation and migration due to dynamical friction. Here we attempt to derive a new center of IKN from the stellar positions of its RGB stars.

\begin{figure}[t]
\includegraphics[width=0.5\textwidth]{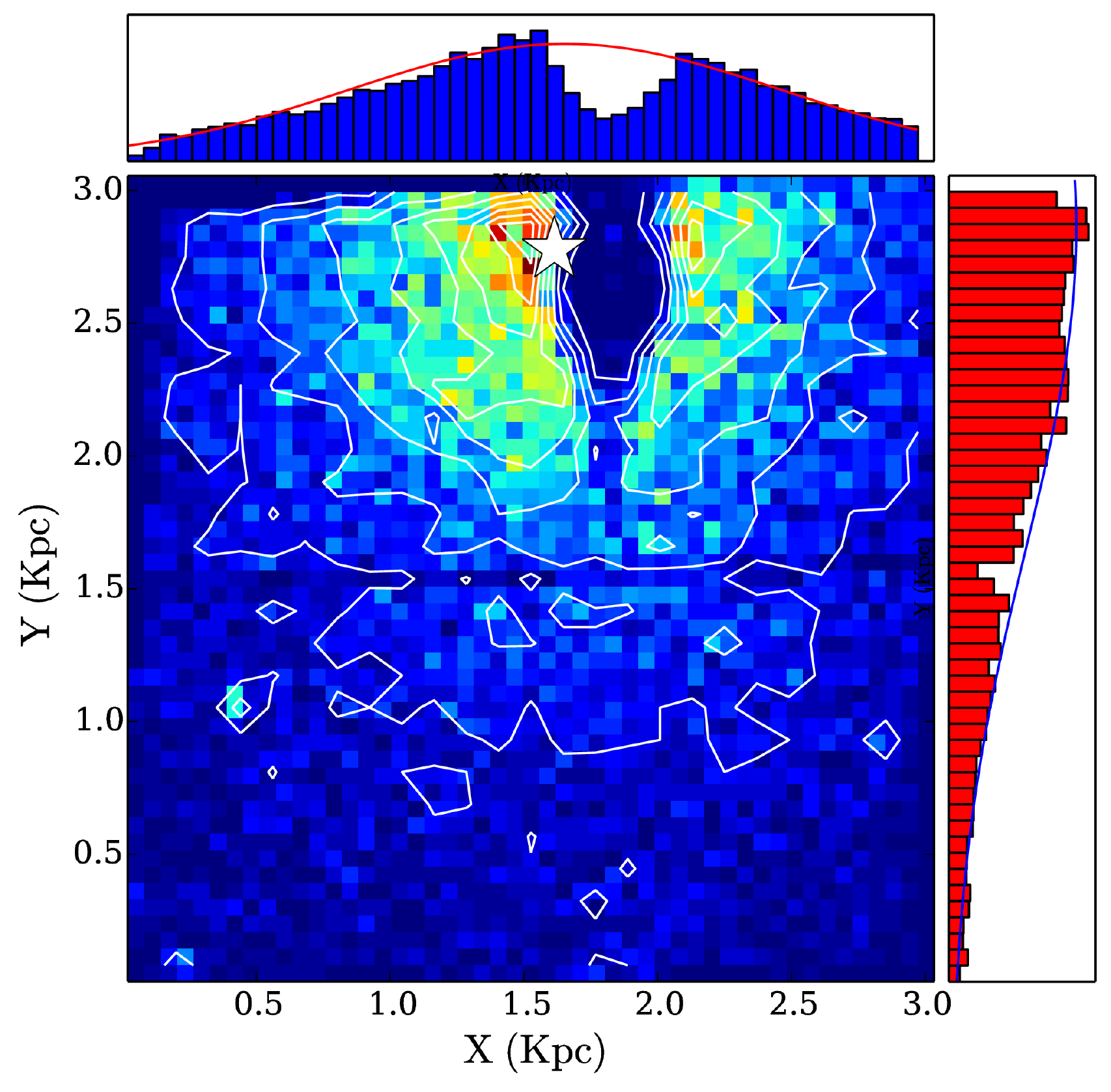}
\caption{Two dimensional number density and histogram projection distributions of the IKN field stars. The histograms have a bin width of 0.345\,kpc and are fitted with a Gaussian function shown with solid lines. The outer countour line in the 2D plot indicates a stellar number density of $17.5\star/{\rm arcsec}^2\ (16\star/{\rm pc}^2$). The white star indicates our best fit position of the IKN centre with an uncertainty of $\pm150$ pc.}\label{Fig:histogram}
\end{figure}
Deriving the radial distribution of the GCs from the IKN centre is a rather complicated task because of the partial coverage of the HST/ACS field and the strong incompleteness due to the bright foreground star. To derive the photometric center of IKN, we used the positions of the field stars detected in the HST/ACS image whose photometric and spatial data was provided to us \cite[by][private communication]{Lianou10}. In Figure \ref{Fig:histogram} we show the two-dimensional number density distribution of the IKN field stars. 
To minimize the impact of the foreground star in deriving the IKN photometric centre, our approach is first to obtain the histogram distributions along the x- and y-axes of the ACS field and then to fit a Gaussian functions to each. Their peaks positions are adopted as the IKN photometric centre. The IKN photometric centre derived using this method is shown with a white star in Figure\,\ref{Fig:histogram} with coordinates of RA$\,=\,10:08:05.8$ and DEC$\,=\,+68:25:20$ and with a total uncertainty of 8.5\arcsec (150\,pc) derived from the fit. We note that the probably the choice of a Gaussian function may not be optimal, but given all other involved uncertainties, it is sufficient for our analysis. Additionally, to obtain a more realistic constraint to the IKN center we fit ellipses to the arcs defined by the constant surface stellar density contours. The 2D surface density of the RGB stars was estimated in a running window with a width of 300\,pixels (15\arcsec). The fit was performed for the arcs that are 300\,pixels inwards from the ACS detector edges and with constant surface density between 3 and 6 stars /arcsec\,$^2$. The results is that the center is at RA$=10:08:24 \pm 45\arcsec$\,; DEC$=+68:24:47 \pm 73\arcsec$\,, PA$=154.4\pm11.1$\,degrees (position angle east of north), ellipticity$=0.216 \pm 0.12$ ($\epsilon=1 - b/a$).

We can use center of IKN to see how the GCs are distributed as a function of the projected galactocentric distance. 
\begin{figure}

\includegraphics[width=0.5\textwidth, bb=10 20 550 390]{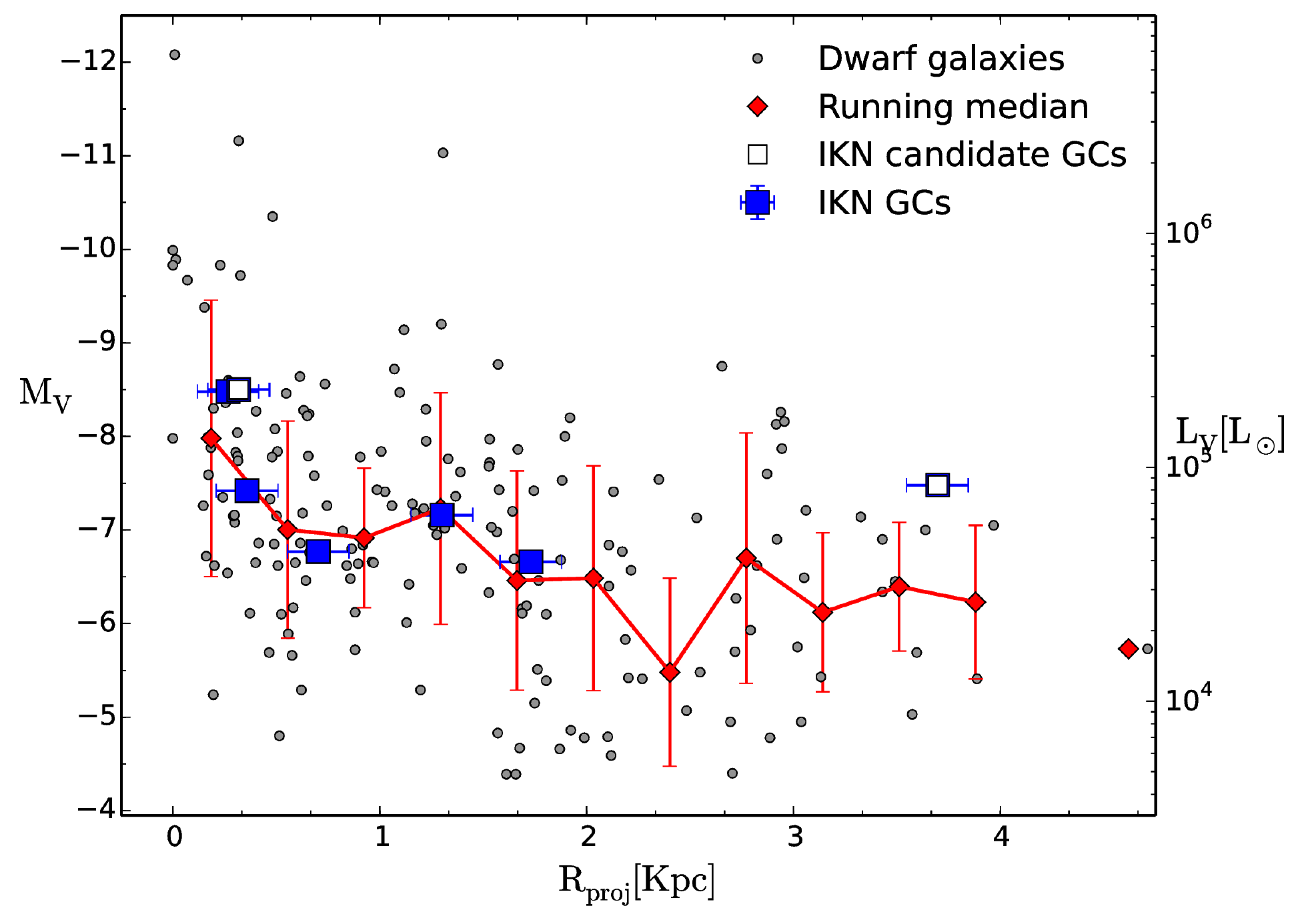}
\caption{Globular cluster luminosity versus projected distance form galaxy photometric center for GCs in dwarf galaxies (small grey circles) and IKN known GCs (large blue squares) and IKN GC candidates (large open squares). The running median (red diamonds) is computed from the entire dwarf galaxies GC sample with a bin size of 0.345\,Kpc, where the y-errorbar is the standard deviation of the number of GCs in each bin.}
\label{Figure:r_proj}
\end{figure}   
The radial distribution of the IKN GCs is shown in Figure\,\ref{Figure:r_proj} with solid (dark blue) squares, where we also show the projected radial distributions of GCs from a larger sample of dwarf galaxies from \cite{Georgiev08,Georgiev09}. In general in Figure\,\ref{Figure:r_proj} is seen that the cluster luminosity (mass) decreases with increasing galactocentric distance. The high mass clusters are closer to the center. To quantify this trend for the combined $R_{\rm proj}$ distributions from all dwarf galaxies, we calculate the running median as a function of $R_{\rm proj}$ with a bin size of 0.345\,kpc (diamonds connected with a solid line). 
Here we confirm an earlier finding that more luminous clusters in dwarf galaxies are typically more centrally located \citep{Georgiev09b}.
Clearly, in Figure\,\ref{Figure:r_proj} is seen that the most massive IKN GC is closest to the derived here photometric center. A likely explanation is that dynamical friction may have ``sorted'' the IKN GCs. However, spectroscopic observations are required to know the local velocity field and whether dynamical friction can be an efficient process for IKN central stellar density.

\subsection{Alignment of IKN GCs and orientation in the M81 group}\label{Section:GCs_alignment}

We also report that the IKN GCs projected positions appear to be linearly aligned. Although this could be also a chance alignment due to low number statistics, the exact spatial and kinematic structure of the IKN GCs needs to be spectroscopically confirmed, because it can have an important implication for interpreting the formation origin of IKN, its SFH and interaction within the M\,81 group of galaxies. Furthermore, a similar linear alignment has been previously reported for the old GCs in NGC\,6822 \citep{Hwang11,Huxor13}.

\begin{figure}

\includegraphics[width=0.5\textwidth, bb=60 20 290 240]{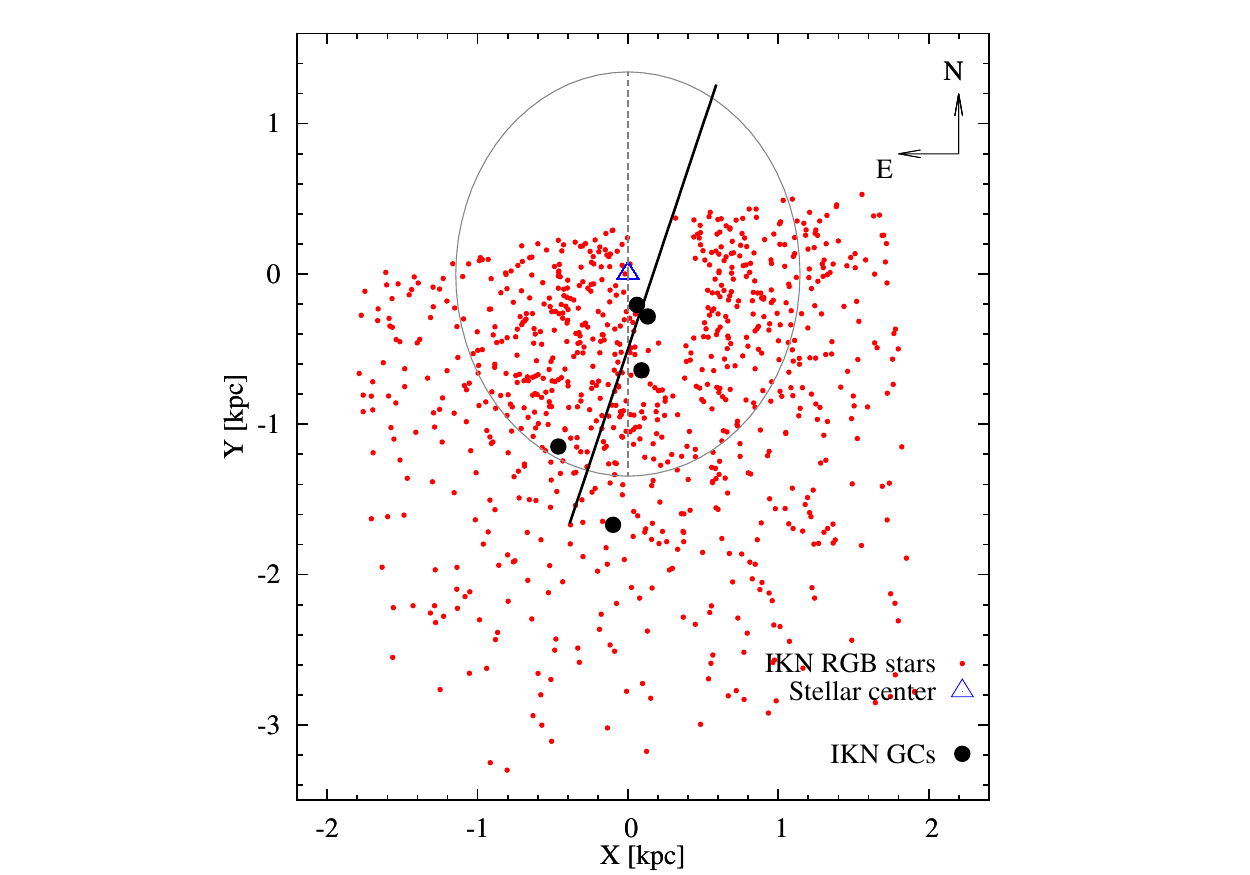}
\includegraphics[width=0.5\textwidth, bb=10 20 500 450]{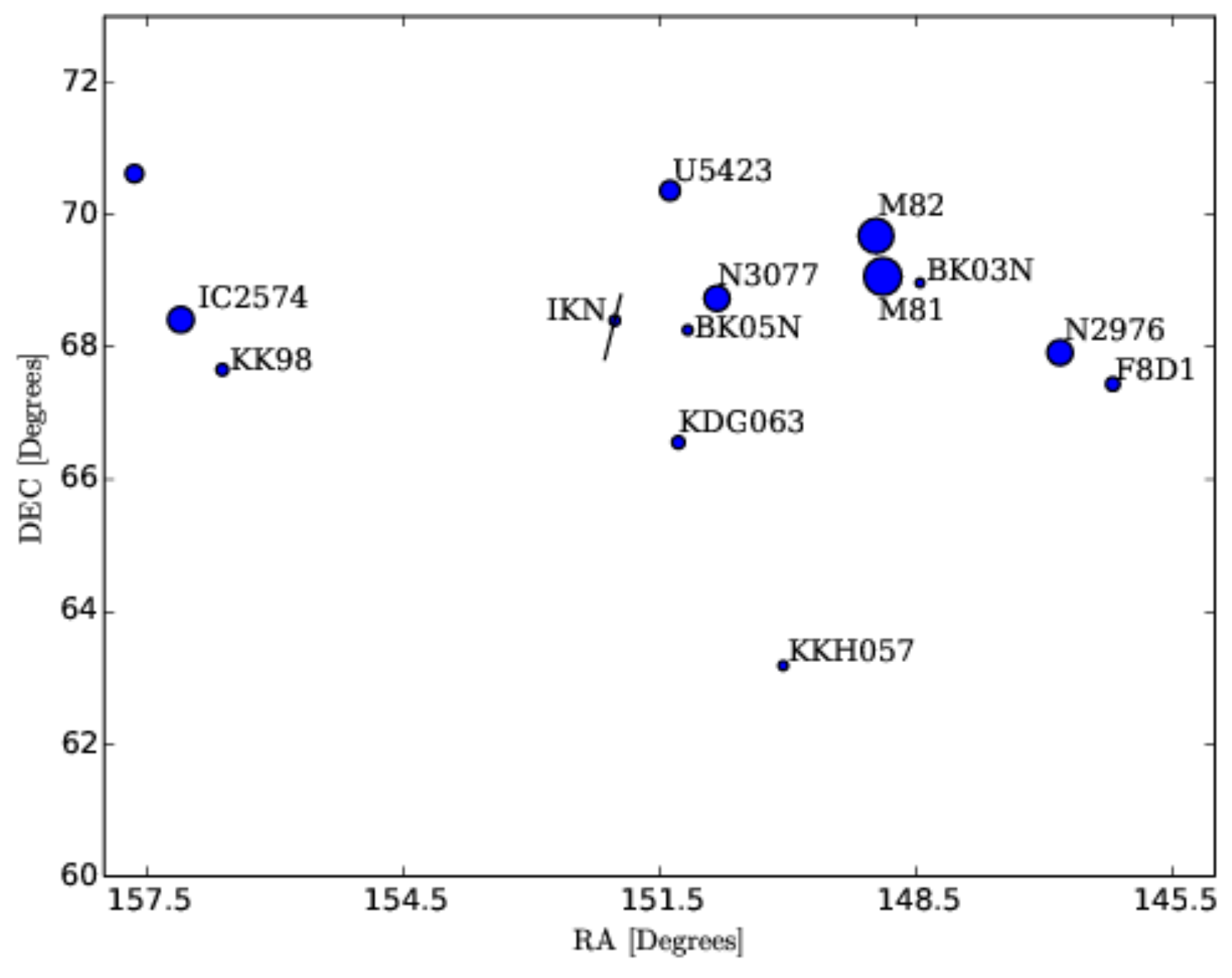}
\caption[RA_DEC_distrib]{{\bf Top:} Spatial distribution of the IKN GCs (large solid circles). The fitted linear regression through their position is shown with solid line, while the IKN major axis and orientation are shown with dashed and light gray line and ellipse, respectively. IKN RGB stars are shown with small (red) dots. {\bf Bottom:} Positions of galaxies associated to the M81 group according to \cite{Karachentsev02}. Circle sizes are proportional to the galaxy absolute B-band luminosity. The line passing through IKN roughly shows the orientation of its GCs from the top panel.
}
\label{Figure:RA_DEC_distrib}
\end{figure}  
The projected linear alignment between the IKN GCs can be already seen in Figure\,\ref{Figure:IKN_HST_Kband}. In Figure\,\ref{Figure:RA_DEC_distrib}\,a) we show the straight line least squares fit through the GCs' positions. For reference, we show with an ellipse the size of IKN's Holmberg diameter, 2.69\,kpc, (measured at $\mu_{B}=25$\,mag) and its ellipticity, $\epsilon=1-b/a=0.18$, retrieved from the HyperLEDA and NED databases. For the center of the ellipse we used the center we obtained in \S\,\ref{Sect:spatial_dist_center}. We note however, that the relative uncertainty in the IKN ellipticity and particularly in the orientation of its major axis reaches nearly $100\%$. This thus prevents us from a more reliable discussion on the alignment between the IKN major axis and the fitted line through the positions of its GCs. 

In the bottom panel of Figure.\,\ref{Figure:RA_DEC_distrib} we compare the orientation of the GCs alignment plane with respect the large scale distribution of galaxies within the M\,81 group. If the GCs alignment is due to their formation in a disk or triggered by interactions with galaxies in the M\,81 group, angular momentum conservation arguments would suggest a correlation with the positions of the nearest galaxies and the larger structure in the group. Figure\,\ref{Figure:RA_DEC_distrib} shows that the GC's projected alignment and the IKN major axis seem to be orthogonal to the larger scale galaxy distribution, i.e. the plane connecting IC\,2547--NGC\,3077--M\,81. However, spectroscopic observations are required in order to properly describe the kinematics of the IKN GCs, whether they are indeed on a (rotating?) disk and what is the orientation of the vector of the angular momentum.

\subsection{Implications for the formation origin of IKN}
         
Hints of the unusual and probably complex star formation history of IKN were first highlighted by the presence of a large number of old globular clusters compared to its total luminosity, i.e. the highest known GC specific frequency of $S_N=126$ \citep{Georgiev09,Georgiev10}. Subsequently, resolved stellar population HST/ACS CMD analysis \citep{Lianou10}, revealed that IKN has a very broad metallicity distribution function. It peaks at a much higher metallicity for the IKN luminosity, unlike other early-type dwarf galaxies in the M\,81 group with similar luminosity \citep{Lianou10}. This led those authors to consider IKN as a galaxy which may have formed as a tidal dwarf galaxy during galaxy-galaxy interactions in the M\,81 group. Recently, \cite{Larsen14} performed a direct spectroscopic comparison between the brightest IKN GC-5 and those of another faint dwarf galaxy - WLM, who showed that in both galaxies the GCs are metal poor, although the IKN-05 GC spectra is of a very low $S/N$. 

Our analysis here further strengthens the idea of a complex SFH for IKN, based on its GCs' ages, metallicity and spatial distributions. Although the photometric uncertainties render large errors in these quantities, the photometric ages and metallicities of the GCs places them among the metal poor field stellar component (cf Fig.\,\ref{Figure:fe_hist}). This suggests that the GCs and the metal poor field stellar component have formed during the same burst of star formation. Our analysis of the field RGB star MDF suggests that there must have been at least two major star formation episodes in the IKN SFH. If galaxy-galaxy interactions in the M\,81 group have triggered bursts of intense star formation, likely galactic pericenter passages were responsible for the multiple peaked SFH. If IKN has formed as a tidal dwarf galaxy, one would also expect that it has a higher for its luminosity metallicity, which is the case. The apparent alignment of the IKN GCs, if real, also would suggest that IKN could be a dSph transformed from a late-type disk dwarf.

The relevant observational evidence for a complex SFH of IKN are: $i)$ on average slightly more metal-rich GCs [Fe/H$]=-1.4$\,dex (cf. Fig.\,\ref{Figure:fe_hist}), than typical for a truly metal-poor GC population of [Fe/H$]<<-1.7$\,dex, e.g. in the Fornax dSph all GCs are with [Fe/H$]<-2.0$\,dex \citep{Larsen14b, Hendricks14}, $ii)$ non-unimodal MDF of the field RGB stars combined with the apparent alignment of the IKN GCs place constraints to the possible scenarios that could have led to the complex IKN SFH. Namely, 1) the IKN GCs should have formed together with the main galaxy stellar population, so that the GCs' and the RGB metallicities and MDFs are matched (Fig.\,\ref{Figure:fe_hist}); 2) this must have happened more than 5-8Gyr ago; 3) the intensity of star formation must be such that it leads to the formation of an unusually large number of GCs or 4) IKN must have lost at least 50\% of its initial mass in order to lead to the high GC specific frequency. Tidal formation origin and/or the tidal interaction triggered star formation scenarios can match these observations. However, if IKN is a primordial dwarf it must contain a significant amount of dark matter, which cannot be estimated for the currently available data. Clearly, a spectroscopic follow up is needed to assess the dynamics of the system.

\section{Summary and conclusions}\label{Section:Conclusions}

We present a new $K-$band photometric analysis of the globular clusters of the IKN dwarf spheroidal using data taken with the NIR detector (LIRIS) installed on the 4.2m William Herschel Telescope. Combined with existing HST/ACS V and I data, we effectively resolve the age-metallicity degeneracy in the $I-K_s$ vs $V-I$ colour space. The new $K_S-$band photometry is tabulated in Table\,\ref{Table:gc_properties1}.

$\bullet$ Using SDSS DR10 archival images we performed a PSF photometric analysis to look for GC candidates in the field around IKN that is not covered by the HST/ACS observations. We found two new GC candidates based on departure form a stellar PSF (cf Fig.\,\ref{Fig: Sharp vs Mag}, \S\,\ref{Section:SDSS_candidates}). One of the candidates (GC-6) is however very close to a bright star which renders its measured parameters highly inaccurate. High spatial and spectral resolution and multi-wavelength observations are required to reveal the nature of these two objects. If one of these GC candidates proves to be another genuine GC, this will make IKN the galaxy which formed GCs with a very high efficiency per unit luminosity with one of the highest specific frequencies known to date,  $S_N=126$.

$\bullet$ To derive GCs' photometric ages, masses and metallicities for all known GCs in the IKN galaxy in \S\,\ref{Section:Photometric_props} we interpolate between \cite{Bruzual03} SSP model tracks (cf Fig.\,\ref{Figure:CCD}). For all GCs we derive mean photometric age which is consistent with the oldest ages in the SSP model grid ($14.29^{+1.01}_{-1.21}$ Gyr, Fig.\,\ref{Figure:MassZ}, Table\,\ref{Table:gc_age_Z_Mass}). The old GC age indicates an ancient burst of star formation. Using the empirically established observational relation between the most massive cluster and the galaxy star formation rate (see \S\,\ref{Sect:PeakSFR}), we find that the burst of star formation led to the formation of $3\times10^5M_\odot$ GC was about $10 \, M_{\odot}/year$ at about 14 Gyr ago (cf Fig.\,\ref{Figure:AgeSFR}, in \S\,\ref{Sect:PeakSFR}). 
 
$\bullet$ We derive GC metallicities in the range $2.0<[$Fe/H$]<-1.2$\,dex (Tab.\,\ref{Table:gc_age_Z_Mass}, \S,\ref{Sect:MDFs}). These are consistent with the metal-poor component of the metallicity distribution function of the field RGB stellar population of IKN reported by \cite{Lianou10}. Our reanalysis of the field RGB stars MDF, data kindly provided to us by \citeauthor{Lianou10}, confirms their finding of a broad MDF. However, we find at least two major bursts of star formation with a small statistical significance, the most recent of which produced the most metal rich stellar population with a peak of [Fe/H$]\simeq-1.0$\,dex (Fig.\,\ref{Figure:fe_hist}). Although the photometric metallicities of the IKN GCs are large, they are more consistent with the metal-poor component. Our mixture model analysis suggests that about 72\% of the IKN stellar population could have formed at the same star-formation epoch as the bulk of the IKN GCs. We note however, that a spectroscopic confirmation is necessary in order to firmly establish the IKN metallicities. For GC-5 such a confirmation does exist and it falls within the photometric uncertainty estimate.

$\bullet$ We find that the more luminous (massive) IKN GCs are preferentially distributed toward the center of IKN (\S\,\ref{Sect:spatial_dist_center}), consistent with earlier observations (cf Fig.\,\ref{Figure:r_proj}). Such mass segregation has been suggested to arise due to dynamical friction arguments, which is efficient in low velocity systems, but strongly depends on the local stellar density. Further spectroscopic observations are required to map the velocity field of IKN and establish whether dynamical friction wold be efficient for such a low stellar surface density galaxy.

$\bullet$ We find that the projected positions of the IKN GCs appear to be aligned. (Fig.\,\ref{Figure:RA_DEC_distrib}, \S\,\ref{Section:GCs_alignment}). This is similar to the observed alignment of the GCs in the Local Group dSph NGC\,6822 \citep{Hwang11,Huxor13}. It remains to be shown spectroscopically whether the IKN GCs have disk kinematics or the apparent distribution is a chance alignment. Knowing this will help for the interpretation of the IKN origin, as a merger of dwarf galaxies will destroy existing disk structures \citep{Bekki08}. The orientation of the plane of alignment of the IKN GCs appears to be very close to the orientation of the IKN major axis, however, the latter is very poorly constrained. Looking at orientation of the GC alignment compared to the large scale distribution of galaxies reveals tentatively that it is orthogonal to the large plane connecting several of the M\,81 group galaxies (see \S\,\ref{Section:GCs_alignment} and Fig.\,\ref{Figure:RA_DEC_distrib} bottom panel).

In summary, the large cluster formation efficiency per galaxy mass, high GC metallicities and positional alignment, point towards a complex SFH for IKN. Whether the   tidal origin scenario for the galaxy indeed holds remains to be cleared through further spectroscopic investigations.

\begin{acknowledgements} 

AT is supported by the DFG Emmy Noether grant Hi 1495/2-1 and the Bonn-Cologne Graduate School of Physics and Astronomy. IG would like to thank for the partial support during the completion of this paper the science department and ``International Research Fellow'' program of ESA-ESTEC in Noordwijk. ACS acknowledges the receipt of CNPq/BJT-A fellowship 400857/2014-6.  The authors would like to thank Sophia Lianou for kindly providing us with their photometric tables and measurements of the IKN field RGB stars metallicities. We are also thankful for the insightful discussions with Søren Larsen. \\
This research has made use of the NASA/IPAC Extragalactic Database (NED) which is operated by the Jet Propulsion Laboratory, California Institute of Technology, under contract with the National Aeronautics and Space Administration. We acknowledge the usage of the HyperLeda database (http://leda.univ-lyon1.fr).

\end{acknowledgements}
           
\bibliographystyle{aa}
\bibliography{references}           
           
\clearpage
\begin{landscape}
\begin{deluxetable}{lllllllllll}
\tabletypesize{\footnotesize}
%\rotate % For landscape mode
%\setlength{\tabcolsep}{0.15cm}
\tablecolumns{11}
\tablewidth{0pc} %%% <--- This is important!!! Otherwise it wont compile!!!!
\tablecaption{Photometric properties of the IKN globular clusters (GCs) and the new candidates (GCCs).
\label{Table:gc_properties1}
}
\tablehead{

\colhead{ID} &
\colhead{DEC} &
\colhead{RA} &
\colhead{$M_V$} &
\colhead{$V-I$} &
\colhead{$K_S$} &
\colhead{$u$} &
\colhead{$g$} &
\colhead{$r$} &
\colhead{$i$} &
\colhead{$z$} \\
\colhead{} &
\colhead{hh:mm:ss} &
\colhead{dd:mm:ss} &
\colhead{mag} &
\colhead{mag} &
\colhead{mag} &
\colhead{mag} &
\colhead{mag} &
\colhead{mag} &
\colhead{mag} &
\colhead{mag} \\
\colhead{(1)} &
\colhead{(2)} &
\colhead{(3)} &
\colhead{(4)} &
\colhead{(5)} &
\colhead{(6)} &
\colhead{(7)} &
\colhead{(8)} &
\colhead{(9)} &
\colhead{(10)} &
\colhead{(11)} 
}
\startdata

GC-1	&	+68:23:36.06	&	10:08:07.12	  &	   $-6.66\pm0.06$  &	   $0.911\pm0.079$   &    $19.225\pm0.089$	&   \nodata		&   $21.955\pm0.097$	&	$21.087\pm0.062$	&   $20.76\pm0.087$	&   $19.876\pm0.125$   \\
GC-2	&	+68:24:04.95	&	10:08:10.80       &        $-7.16\pm0.07$  &	   $0.994\pm0.078$   &    $18.622\pm0.054$	&   \nodata		&   $21.321\pm0.054$	&	$20.648\pm0.052$	&   $20.372\pm0.068$	&   \nodata	   \\
GC-3	&	+68:24:33.16	&	10:08:05.27       &        $-6.77\pm0.06$  &	   $1.085\pm0.092$   &    $18.891\pm0.071$	&   \nodata		&    \nodata	   	&     \nodata	  		 &     \nodata	   &     \nodata	   \\
GC-4	&	+68:24:53.07	&	10:08:04.80       &        $-7.42\pm0.06$  &	   $0.936\pm0.077$   &    $18.321\pm0.043$	&   \nodata		&   $20.997\pm0.052$	&	$20.424\pm0.044$	&   $20.098\pm0.053$	&   $19.941\pm0.135$   \\
GC-5	&	+68:24:57.35	&	10:08:05.52       &        $-8.48\pm0.06$  &	   $0.906\pm0.077$   &    $17.293\pm0.019$	&$21.435\pm0.174$	&   $19.911\pm0.025$	&	$19.312\pm0.026$	&   $19.097\pm0.044$	&   $18.769\pm0.057$   \\
GCC-6	&	+68:25:29.13	&	10:08:05.16       &	   $-8.50\pm0.04$*	   &	\nodata	     &		\nodata	   	&$20.895\pm0.124$	&   $19.675\pm0.035$	&	$19.045\pm0.03$		 &	$18.651\pm0.049$   &	 $18.716\pm0.086$   \\
GCC-7	&	+68:28:36.72	&	10:08:08.85       &        $-7.48\pm0.04$*	   &	\nodata	     &		\nodata	   	&    \nodata		&   $20.766\pm0.038$	&	$20.006\pm0.08$		 &	$19.895\pm0.051$   &	 $19.762\pm0.107$   \\

\enddata

\vspace{-.5cm}
\tablecomments{Absolute magnitude in column (4) is calculated using the distance modulus of $m-M=27.79\pm0.02$\,mag adopted in the paper. The SDSS $u,g,r,i,z$ are the PSF magnitudes derived in this work. Johnsons/Cousins $V,I$ magnitudes are from \cite{Georgiev09}.}

\end{deluxetable}
\end{landscape}

\end{document}